\documentclass[final,5p,times,fleqn]{elsarticle}
\usepackage{graphicx}% Include figure files
\usepackage{dcolumn}% Align table columns on decimal point
\usepackage{bm}% bold math
\usepackage{bm}% bold math
\usepackage{mciteplus}
\usepackage{amsmath}
\usepackage{amsfonts,amssymb}
\usepackage{color}
\usepackage{cancel}
\usepackage{enumerate,soul}
\usepackage{amssymb}

\begin{document}
\begin{frontmatter}
\title{On Emergent Gravity, Ungravity and $\Lambda$}% Force line breaks with \\
\author{Luis Rey Diaz-Barron$^{a}$}
\ead{lrdiaz@ipn.mx}
\author{M. Sabido$^{b}$}
\ead{msabido@fisica.ugto.mx}
\address{  $^a$Instituto Polit\'ecnico Nacional, Unidad Profesional Interdisciplinaria
	de Ingenier\'ia Campus Guanajuato.\\ C.P. 36275, Silao de la Victoria,
	M\'exico.\\
$^b$ Departamento  de F\'{\i}sica de la Universidad de Guanajuato,\\
 A.P. E-143, C.P. 37150, Le\'on, Guanajuato, M\'exico.\\

 }%

\begin{abstract}
In this paper we study the {\it``ungravity"} modifications to the Friedmann equations. By using the first law of thermodynamics and  the modified entropy-area relationship derived from the {\it``ungravity"} contributions to the Schwarzschild black hole, we obtain modifications to the Friedmann equation that in the late time regime gives an effective cosmological constant. Therefore, this simple model can provide an {\it ``ungravity"} origin to the cosmological constant $\Lambda$.
\end{abstract}

\end{frontmatter}

%%%%%%%%%%%%%%%%%%%%%%%%%%%%%%%
\section{Introduction} \label{Int}
%%%%%%%%%%%%%%%%%%%%%%%%%%%%%%%
The discovery of the late time acceleration of the Universe has deeply impacted our understanding of fundamental physics.  If we propose a solution in the context of general relativity (GR) we must asume the existence of a primordial energy density or a new type of matter (not contained in the SM) with the  property of having a negative pressure, currently a cosmological constant $\Lambda$ seems to be favored. Although  observations are compatible with a nearly vanishing but positive  cosmological constant, there are inconsistencies with traditional quantum field theory \cite{weinberg,lambda}. 
%The current explanation of this phenomenon is inconsistent with the standard model of particle physics (SM) and with General Relativity (GR). 
The problems related to the cosmological constant have been addressed by different approaches \cite{polchinski}.
There is no known mechanism that guarantees a zero or nearly zero value for $\Lambda$ in a stable vacuum, or explain the similar value  between the associated energy density of  $\Lambda$ and the energy density of present day matter.  But the biggest problem is the 120 orders of magnitude discrepancy between the predicted vacuum energy density derived from QFT  and the energy density associated to $\Lambda$.
Of course  we entertain the possibility that the problems related to $\Lambda$ and the current acceleration are  consequence of the poor understanding of gravity and therefore a reformulation of gravity is warranted.

A recent approach to understand the incompatibility of gravity with quantum mechanics is to consider the gravitational interaction  as an emergent phenomenon. In \cite{Jacobson:1995ab}, starting with an entropy proportional to the area  Einstein's equations are derived, proving that one can consider GR as an entropic force. Interest on the entropic origin of gravity was rekindled  by Verlinde  in \cite{Verlinde:2010hp}, where he explores an entropic origin to Newtonian gravity, proposing that gravity is an entropic force, similar to the emergent forces that are present in polymers. Since in this formulation Newtonian gravity has an entropic origin, by introducing changes to the Bekenstein-Hawking entropy-area relation  one can induce modifications toe Newtonian gravity (see \cite{isaac} and references therein).
 In a more recent paper \cite{Verlinde2}, Verlinde claims that the dark matter  and dark energy problems can be simultaneously solved in the emergent formulation of gravity. 
The dark matter predictions for this theory have been put to test in several works \cite{niz,test2}. Even if Verlinde's proposal, is not a rigorous  formulation, it has some intriguing ideas that warrant a more detailed exploration. In particular the connection to the dark matter and dark energy problems gives a new approach to study dark sector of the Universe.

In the last decade, the presence of {\it ``un-particle"} degrees of freedom in low energy physics  was extensively explored in the context of particle physics, these effect are originated from a UV scale invariant sector  that couples to the SM. Scale invariant QFT's  represent massless particles, if we extend the notion of scale invariance to include a new kind of {\it ``particle''} will have a continuos mass spectrum and therefore no definite mass, for this reason this extension is known as {\it ``un-particle"} sector \cite{Georgi:2007ek}. The effects of the {\it ``un-particle"} sector are easily calculated and can be probed in current accelerators, unfortunately the search for these effects has been negative. 
Although the search of the {\it ``un-particle"} effects in the standard model has yielded negative results one can explore them in the context of gravity. There have been attempts to study the testability of {\it ``un-particle"} effect in connection to Newtonian gravity, the analysis is based on scalar modifications \cite{Goldberg:2008zz} to the gravitational potential. {Also in the cosmological scenario,   {\it ``un-particle"} effects have been studied} in order to impose a lower bound on the scale of the interaction of the  {\it ``un-particle"} sector  \cite{dai}. This approaches introduce an {\it ``un-particle"} for a scalar field or to the standard model sector and not to a graviton, therefore is not extension to GR (we do have to clarify that in \cite{Georgi:2007ek} they are working with extension to Newtonian gravity so the modifications can be considered extensions to gravity). 
If we want to consider {\it ``un-particle"} extension to GR we need to assume the existence of {\it ``un-gravitons"}. In \cite{ungravity}, the authors find an effective action that that allows the study of gravitational effects beyond the weak field approximation. In  \cite{unbh}, using an effective action that incorporates {\it ``un-particle"} effects to GR, the authors study the ungravity  effects to Schwarzschild black hole and derive its temperature and entropy. 

With these ideas in mind we  explore the late time behavior of the Friedmann-Robertson-Walker (FRW) universe in the context of entropic gravity introducing the effects of the {\it ``ungravity"} sector. For this we will calculate the modification to Friedmann equations and analyze it late time behavior. In order to calculate the modifications to the Friedmann equations, we will use the first law of thermodynamics \cite{rongen, cai} with the modified entropy-area relationship. We will use the entropy-area relationship derived from the {\it ``ungravity"} effective action for the Schwarzschild black hole. Finally analyzing the late time behavior, we find an effective cosmological constant that is  related to the {\it ``ungravity"} parameters. 

The paper is organized as follows. In section \ref{ungrav}, we briefly review the ungravity modification to gravity and the new entropy area relationship and  derive the modified Friedmann equations. Section \ref{final}, is devoted for discussion and final remarks.

%%%%%%%%%%%%%%%%%%%%%%%%%%%%%%%
\section{The Ungravity  Friedmann Equations} \label{ungrav}
%%%%%%%%%%%%%%%%%%%%%%%%%%%%%%%
As is well known the Friedmann equations together with the continuity equation, is all that is needed to study the dynamics of the universe. Using the Clausius equation and a linear relationship between the entropy and the area $S\approx A$, one can derive Friedmann equations \cite{rongen}. Furthermore, to study modifications to the Friedmann equations, we can start from a modified entropy-area relationship and derived the modified equations \cite{cai}. In this approach the new {\it ``physics''} is encoded in the entropy-area expression.

If we are to explore the effects of the {\it ``ungravity} sector to cosmology, the we must use the entropy-area relationship that includes the effects of the {\it ``ungravity"} degrees of freedom \cite{ungravity}.   Using this effective action to the case of the Schwarzchild black hole, the {\it ``ungravity} contributions to the temperature and entropy of the black-hole are derived \cite{unbh}.
The temperature for the Schwarzschild black hole after incorporating the {\it ``ungravity} effects  is 
\begin{equation}
T_U=\frac{\hbar c}{4\pi k_B\tilde r_h}\left[1+\frac{2(2d_U-1)\Gamma_U}{1+\Gamma_U\left(\frac{R}{\tilde r_h}\right)^{2d_U-2}}\left(\frac{R}{\tilde r_h}\right)^{2d_U-2}\right],
\end{equation}
where $\Gamma_U$ is defined as
\begin{equation}
\Gamma_U=\frac{2}{\pi^{2d_U-1}}\frac{\Gamma(d_U-1/2)\Gamma(d_U+1/2)}{\Gamma(2d_U)},
\end{equation}
and $R$ is
\begin{equation}
R=\frac{1}{\lambda_U}\left(\frac{M_{\rm{Pl}}}{M_U}\right)^{1/(d_U-1)}.
\end{equation}
The constants, $M_U$  is the {\it ``ungravity"} coupling constant and is related to the interaction between the {\it ``un-graviton"} and the usual matter, and $\lambda_U$ is  the critical energy scale at which the scale invariant properties of {\it ``un-graviton"} emerge. Finally, $d_U$ is the scaling parameter that labels the continuos mass spectrum of the {\it ``un-graviton"} and can take values $1<d_U<2$. The thermodynamic energy of the system as a function of the horizon is
\begin{equation}
U(\tilde{r}_h)=\tilde{r}_h\frac{c^4}{2G}\left(\frac{1}{1+\Gamma_U\left(\frac{R}{\tilde{r}_h} \right)^{2d_u-2}}\right),
\end{equation}
from which we obtain
\begin{equation}
dS_U=d(A)\frac{k_Bc^3}{4\hbar G}\left(\frac{1}{1+\Gamma_U\left(\frac{R}{\tilde r_h}\right)^{2d_U-2}}\right).\label{entropy}
\end{equation}
Finally integrating the previous equation, we get the {\it``ungravity"} entropy
\begin{equation}
\label{unentropy}
S=\frac{k_Bc^3}{4 \hbar G}\frac{ (2\sqrt{\pi}R)^{2-2d_U}}{d_U\Gamma_U}A^{d_U},
\end{equation}
this is the entropy-area relationship that includes the {\it ``ungravity"} effects, we have defined $A=4\pi\tilde{r}^2_h$ in the previous equation. This will be the starting point to derive the {\it ``ungravity"} Friedmann equations.
This new entropy-area relationship was used to obtain the {\it ``ungravity"} modifications to Newtonian gravity \cite{nicolini}, the corrections to Newtonian force. Although the corrections are small, the correction derived using the entropic approach are consistent with the corrections by only considering the {\it ``un-particle"} contribution \cite{gaete}. This gives confidence that the results derived from an entropic formulation of gravity are encoded in the modified entropy-area relationship.

To obtain the {\it ``ungravity"} Friedmann equations we will follow the approach in \cite{rongen}, where  the Friedmann equations are derived using the Clausius equation and the  entropy-area relationship evaluated at the apparent horizon.

Let us proceed with the derivation of the Friedmann equation, for this  we start with the FRW metric, 
\begin{equation}
ds^{2}=-dt^{2}+a^{2}(t)\left( \frac{dr^{2}}{1-\kappa r^{2}}+r^{2}d{\Omega}^{2}\right),
\end{equation}
comparing with
$
ds^{2}=h_{ab}dx^{a}dx^{b} +\tilde{r}^{2}d\Omega^{2},
$
we can identify $h_{ab}$.
Following the procedure in \cite{rongen}, we introduce the work density $W$ and the  energy-supply vector $\Psi$
\begin{equation}
W=-\frac{1}{2}T^{ab}h_{ab}, \qquad \Psi_a=T^b_a\partial_b\tilde r+W\partial_a\tilde r,
\end{equation}
where $\tilde{r}=a(t)r$ and $T_{ab}$ is the projection of $T_{\mu\nu}$ in the normal direction of the 2-sphere.  
{Using the  energy-momentum tensor for a perfect fluid, the work density and the energy supply vector are}
\begin{eqnarray}
W&=&\frac{1}{2}(\rho-P),\\
\Psi_{a}&=&\left( -\frac{1}{2}(\rho+P)H\tilde{r},\frac{1}{2}(\rho+P)a\right).\nonumber
\end{eqnarray}
The amount of energy $\delta Q$ crossing the apparent horizon during the time interval $dt$ is given  by
\begin{equation}
\delta Q=-A\Psi=A(\rho +p)H\tilde r_hdt, 
\end{equation}
where $A=4\pi\tilde{r}^2_h$ is the area of the apparent horizon and $\tilde{r}_{h}$ is the radius of the apparent horizon 
\begin{equation}
\tilde{r}_h=\frac{1}{\sqrt{H^2+\kappa/a^2}},
\end{equation}
%and is obtained from $h^{ab}\partial_{a}\tilde{r}\partial_{b}\tilde{r}=0$, a direct calculation gives 
the apparent horizon is obtained from $h^{ab}\partial_{a}\tilde{r}\partial_{b}\tilde{r}=0$. 

From {this point forward} we will use the area-entropy relationship defined at the apparent horizon  this  area-entropy relationship is analogous to one we get at the black hole horizon. With this hypothesis, in \cite{rongen} the authors the derive the usual Friedman equations.%and will use  the modified entropy-area relationship Eq.(\ref{entropy}), to derive the modified Friedmann equation.

Using the Clausius relation $\delta Q=T_UdS$, and the continuity equation $\dot\rho=-3H(\rho+p)$,  we get
\begin{flalign}
\label{ClausR}
\frac{8\pi G}{3}\frac{\partial \rho}{\partial t}&=\frac{c^4}{2}\left[1+\frac{2(2d_U-1)}{1+\Gamma_U(R^2(4\pi/A))^{d_U-1}}\Gamma_U\left(R^2\frac{4\pi}{A}\right)^{d_U-1}\right]\nonumber\\ 
&\times\left[\frac{1}{1+\Gamma_U(R^2(4\pi/A))^{d_U-1}}\right]\frac{d(4\pi/A)}{dt}.
\end{flalign}
Integrating both sides of the equation we get
\begin{flalign}\label{hyp_fried}
&\frac{8\pi G}{3}\rho=\frac{c^4v}{2\Gamma_u^\alpha R^2} {}F^{(2)}_1\left(1,\alpha;1+\alpha;-v^{1/\alpha}\right) \nonumber\\
&-\frac{c^4(2d_U-1)}{\Gamma_u^\alpha R^2}\frac{\alpha v}{(1+v^{1/\alpha})}\\
&+\frac{\alpha c^4(2d_U-1)\nu}{\Gamma_u^\alpha R^2}{}F^{(2)}_1\left(1,\alpha;1+\alpha;-v^{1/\alpha}\right),\nonumber
\end{flalign}
where $\alpha=1/(d_U-1)$ and $v= 4\pi\Gamma_U^\alpha R^2A^{-1}$.  $F^{(2)}_1(a,b;c;z)$ is the hypergeometric function, for $|z|<1$ is defined by the series expansion
\begin{equation}
F^{(2)}_1(a,b;c;z)=\sum^\infty_{n=0}\frac{\Gamma(a+n)\Gamma(b+n)\Gamma(c)}{\Gamma(a)\Gamma(b)\Gamma(c+n)\Gamma(n+1)}z^n.\nonumber
\end{equation}
To obtain the modified Friedmann equation, we use the area of apparent horizon $A=4\pi/(H^2+k/a^2)$ to finally arrive at
\begin{flalign}\label{unfried}
&\frac{8\pi G}{3}\rho= -\frac{c^4(2d_U-1)}{d_U-1}\frac{H^2+\frac{\kappa}{a^2}}{1+\Gamma_U\left[R^2\left(H^2+\frac{\kappa}{a^2}\right)\right]^{(d_U-1)}}\quad\\
&+\frac{c^4(5d_U-3)}{2(d_U-1)} \sum^\infty_{n=0}\frac{(-1)^n\Gamma_U^nR^{2n(d_U-1)}}{(d_U-1)(n+\frac{1}{d_U-1})}\left(H^2+\frac{\kappa}{a^2}\right)^{n(d_U-1)+1}.\nonumber 
\end{flalign}
{To study the dynamics of the Universe we simply need to introduce the appropriate matter, (i.e, radiation, dust, scalar field, etc.) and  solve the resulting equations. As we are interested in a more generic result  that does not depend on the particular type for the matter density and pressure.
As Eq.(\ref{unfried}) is very complicated, we will make some assumption and approximations.} \\
In gravity, we know there is an holographic bound to the entropy that implies that at most the entropy for the gravitational interaction must be proportional to the area. This bound does not apply for the {\it ``ungravity"} sector, we can see from Eq.(\ref{unentropy}) the entropy $S\sim A^{d_U}$. 

We will take  $d_U=3/2$, as for this case the entropy will scale with the volume. Entropy terms that scale on the volume are related to nongravitational degrees of freedom  and are not present in GR (it is worth mentioning that volumetric corrections to the entropy of black holes have been derived in loop quantum gravity \cite{lqg}). Furthermore in \cite{Verlinde2} it is argued that volumetric contributions to the entropy term can be related to dark energy. After taking $d_U=3/2$ we get
\begin{align}
\label{FE3/2}
\frac{8\pi G}{3}\rho=&\frac{9c^4}{2}\sum^\infty_{n=0}\frac{2(-1)^n\pi^{-2n}R^n}{n+2}\left(H^2+\frac{\kappa}{a^2}\right)^{\frac{n}{2}+1}\nonumber\\
&-4c^4\frac{H^2+\frac{\kappa}{a^2}}{1+\pi^{-2}\left[ R^2\left(H^2+\frac{\kappa}{a^2}\right)\right]^{1/2}}.
\end{align}
This modified Friedmann equation was originally derived in terms of the hypergeometric function, therefore the convergence of r.h.s. of Eq.(\ref{hyp_fried}) is guaranteed for $\vert v^{d_U-1}\vert <1$, therefore  $\Gamma_U R^{2(d_U-1)}(H^2+k/a^2)^{d_U-1}<1$.

For $d_U=3/2$, the constraint takes the form  $\pi^{-2} R(H^2+k/a^2)^{1/2}<1$, from here on the units we are employing we have $c=1$.
In this approximation, to leading order we get 

\begin{equation}
\frac{8\pi G}{3}\rho= \frac{9}{2} \left(\left(H^2+\frac{\kappa}{a^2}\right)-\frac{2 R \left(H^2+\frac{\kappa}{a^2}\right)^{3/2}}{3 \pi ^2}\right)
 -4 \left(H^2+\frac{\kappa}{a^2}\right).
\end{equation}
Now we solve  for $\left(H^2+\frac{\kappa}{a^2}\right)$ and get
\begin{equation}
H^2+\frac{\kappa}{a^2}=\frac{\pi ^2(\pi^2+\sqrt[3]{M_2+M_3})}{108 R^2}-\frac{M_1}{8748 \pi ^2 R^2 \sqrt[3]{M_2+M_3}},
\end{equation}
where $M_1, M_2, M_3$ are defined as follows
\begin{eqnarray}
M_1&=&93312 \pi ^5 g \rho  R^2-81 \pi ^8,\quad\\
M_2&=&497664 g^2 \rho ^2 R^4-1728 \pi ^3 g \rho  R^2+\pi ^6\nonumber\\
M_3&=&13824 \sqrt{1296 g^4 \rho ^4 R^8-\pi ^3 g^3 \rho ^3 R^6}.\nonumber
\end{eqnarray}
As we are interested in the late time evolution, we analyze the limit $t\to\infty$, this is equivalent to the limit limit $\rho\to0$, 
\begin{equation}
H^2+\frac{\kappa}{a^2}=\frac{\pi ^4}{36  R^2 },
\end{equation}
from this equation in this limit, we can define an effective cosmological constant. 

Finally the cosmological constant in terms on the {\it ``ungravity''} parameters is 
\begin{equation}\label{lambda}
\Lambda_{eff}\sim \lambda_U^2\left(\frac{M_U}{M_{pl}} \right)^{\frac{2}{d_u-1}}.
\end{equation}

%%%%%%%%%%%%%%%%%%%%%%%%%%
%%%%%%%%%%%%%%%%%%%%%%%%%%
\section{Discussion and Final Remarks} \label{final}
%%%%%%%%%%%%%%%%%%%%%%%%%%
%The introduction of {\it``un-particle"} effects on gravity where studied in \cite{Goldberg:2008zz}, where the authors consider Newtonian gravity and therefore the  {\it``un-particle"} effects are derived from scalar modifications. The bounds they find are at the scale at which Newtonian gravity is valid.

In this work, we consider the {\it``un-particle"} effects in the cosmological scenario, this is done in order to answer the question, is {\it``un-particle"} physics relevant at the cosmological scale?. The result of the previous section point to a positive answer. In particular it gives insight on a possible origin to the cosmological constant from the  {\it ``ungravity''}  sector.  So let us take seriously this possibility and consider that the cosmological constant is originated from the {\it ``ungravity''} sector. 

We can impose the current value of $\Lambda$ and find a relationship between $M_U$ and $\lambda_U$.
\begin{figure}[htbp] %  figure placement: here, top, bottom, or page
   \centering
   \includegraphics[width=2.5in]{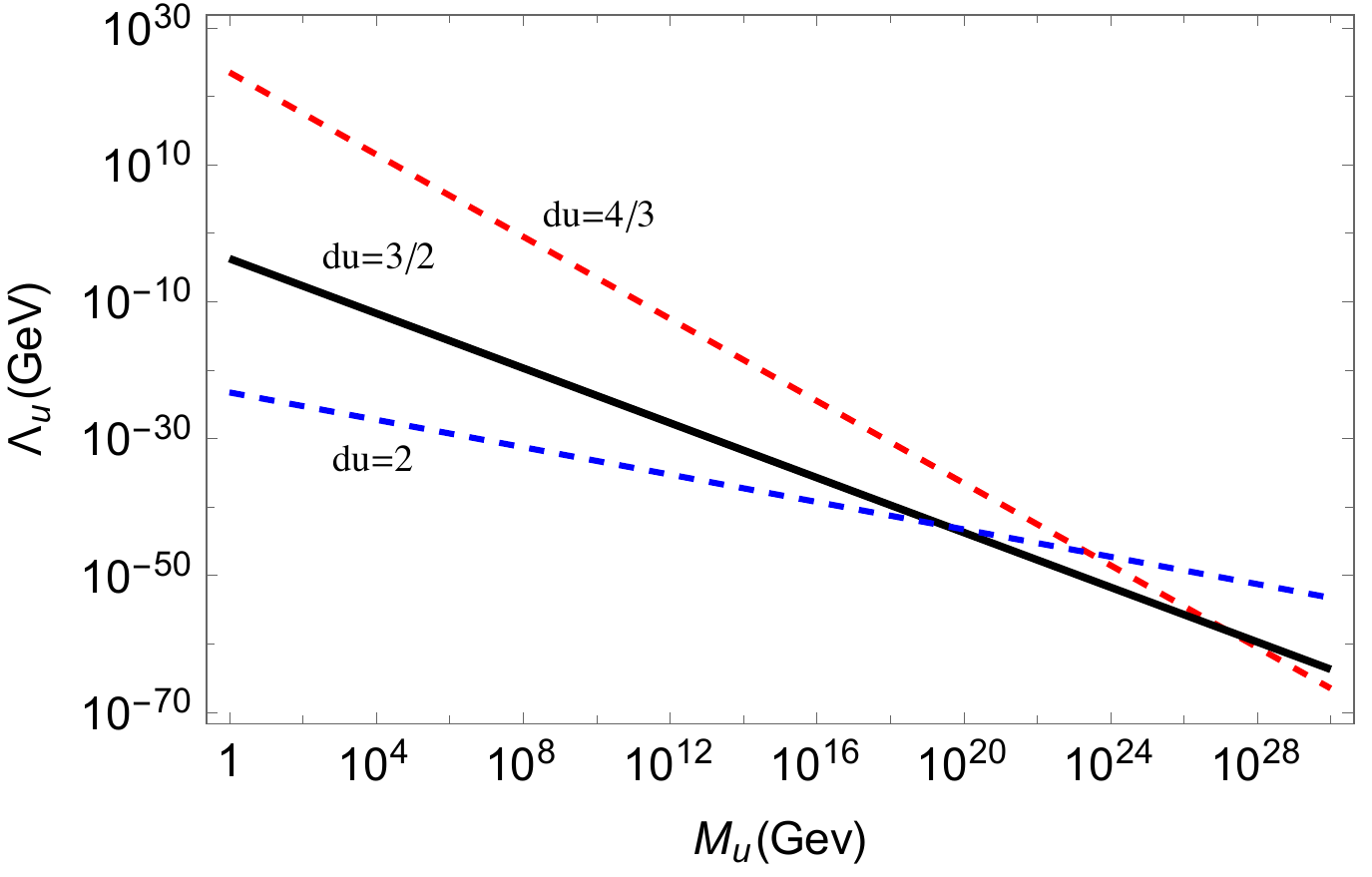} 
   \caption{Valid region of parameters $\lambda_U$ and $M_U$ for different values of $d_U$ in order to have the values for the cosmological constant.}
   \label{figura}
\end{figure}
In fig.(\ref{figura}), the lines represent the values of $\lambda_U$ and $M_U$ that give the correct value for the cosmological constant $\Lambda$. The value $d_U=2$ does not represent a valid {\it ``ungravity''} theory and is presented to show the behavior for the upper value of $d_U$.   For the values of $d_U$ considered in fig.(\ref{figura}) we can rewrite the effective cosmological constant as $\Lambda_{eff}\sim\frac{1}{R^2}$. Then for the case $d_U=\frac{3}{2}$ we have can conclude that the degrees of freedom give a volumetric contribution to the entropy and an effective cosmological constant.

Now we can ask, is Eq.(\ref{lambda}) is valid for other values of $d_U$? If the answer is positive, then this result is more or less general for the {\it``ungravity"} sector, if the answer is negative, at least it is consistent with the fact the volumetric contributions to the entropy are related to the dark energy sector.

For  $d_U=\frac{4}{3}$ Eq.(\ref{lambda}) holds, but for arbitrary values  of $d_U$ an analytical solution for $\Lambda_{eff}$ in terms of $M_U$ and $\lambda_U$ can not be found. Nonetheless, if we consider that for $d_U=\frac{3}{2}+\epsilon$ with $\epsilon$ a small number (we now it holds for $d_U=\frac{4}{3}$) the  functional behavior of $\Lambda_{eff}$ as a function of $R$ is similar, therefore we can argue that the cosmological constant from the {\it ``ungravity''} sector is
\begin{equation}
\Lambda_{eff}\sim\frac{1}{R^2}.
\end{equation}
Of course there are some several factor that we mus consider, first of all $M_U$ and $\lambda_U$ are not necessarily the same that appear in the {\it ``un-particle"} extensions  of the standard model or  the {\it ``un-particle"} scalar extensions  to Newtonian gravity and we have we have to  be careful as we are lacking a fully consistent formulation of {\it ``un-particle"} physics and the problems are inherited by the {\it``ungravity''} theory.

Furthermore, it seems that if we want to attribute the  dark energy sector of the Universe to the gravitational interaction, we need gravitational degrees of freedom whose entropy scales as $S\sim A^m$ with $m>1$. Unfortunately this necessary implies a violation of the holographic principle, this is rather 
reasonable principle  and systems dominated by gravity are 
expected to  fulfill it \cite{holographic}. Therefore on way out of this conundrum is to consider the {\it``ungravity''}  sector, this sector can violate the holographic principle an has an entropy that scales as $S\sim A^{d_U}$ with $1<d_U<2$.
With this in mind we can argue that in the context of  an {\it``ungravity''} theory, we can obtain an effective cosmological constant that depends on the  {\it``ungravity''}  coupling constant  $M_U$  and the critical energy scale $\lambda_U$ that is the scale were invariant properties of {\it ``un-graviton"} emerge.

\section*{Acknowledgements}
This work is supported by CONACYT grants 257919, 258982. M. S. is supported by CIIC 28/2018.

\end{document}